\documentclass[paper,prd,reprint,preprintnumbers,nofootinbib,notitlepage,showpacs,showkeys]{revtex4-1}
\usepackage{latexsym,graphicx,amssymb,amsmath,mathrsfs} 
\usepackage{slashed,mathbbol}
\usepackage[utf8]{inputenc}

\DeclareSymbolFont{bbold}{U}{bbold}{m}{n}
\DeclareSymbolFontAlphabet{\mathbbold}{bbold}

\newcommand{\be}{\begin{equation}}      
\newcommand{\ee}{\end{equation}}      
\newcommand{\bea}{\begin{eqnarray}}      
\newcommand{\eea}{\end{eqnarray}}    
\newcommand{\Tr}{\,\textrm{Tr}\,}
\newcommand{\ife}{\,\textrm{if}\,}
\newcommand{\orr}{\,\textrm{or}\,}

\newcommand{\ch}{\,\textrm{ch}\,} 
\newcommand{\els}{\,\textrm{else}\,} 
\newcommand{\mix}{\,\textrm{mix}\,} 
\newcommand{\ns}{\,\textrm{ns}\,} 
\newcommand{\s}{\,\textrm{s}\,} 
\newcommand{\MeV}{\,\textrm{MeV}\,} 
\newcommand{\GeV}{\,\textrm{GeV}\,}

\makeatletter
\renewcommand\appendix{\par
\setcounter{section}{0}%   
\setcounter{subsection}{0}% 
\gdef\thesection{\appendixname\space\@Alph\c@section}}

\long\def\unmarkedfootnote#1{{\long\def\@makefntext##1{##1}\footnotetext{#1}}}
\makeatother

\begin{document} 

\title{Field dependence of the Yukawa coupling in the three flavor quark--meson model} 
\author{G. Fej\H{o}s}
\email{gergely.fejos@ttk.elte.hu}
\author{A. Patk\'os}
\email{patkos@galaxy.elte.hu}
\affiliation{Institute of Physics, E\"otv\"os University, 1117 Budapest, Hungary}

\begin{abstract}
{We investigate the renormalization group flow of the field-dependent Yukawa coupling in the framework of the three flavor quark--meson model. In a conventional perturbative calculation, given that the field rescaling is trivial, the Yukawa coupling does not get renormalized at the one-loop level if it is coupled to an equal number of scalar and pseudoscalar fields. Its field-dependent version, however, does flow with respect to the scale. Using the functional renormalization group technique, we show that it is highly nontrivial how to extract the actual flow of the Yukawa coupling as there are several new chirally invariant operators that get generated by quantum fluctuations in the effective action, which need to be distinguished from that of the Yukawa interaction.}
\end{abstract}

\maketitle

\section{Introduction}
One of the merits of the functional renormalization group (FRG) technique is that it allows for calculating the flows of n-point functions nonperturbatively. Via the FRG one has the freedom to evaluate the scale dependence of them in nonzero background field configurations, which is thought to be essential once spontaneous symmetry breaking occurs \cite{berges00,gies12,dupuis20}. For the sake of an example, in scalar $(\phi)$ theories, quantum corrections to the wave function renormalization ($Z$) typically vanish at the one-loop level, but using the FRG one gets a nonzero contribution once one generalizes the corresponding kinetic term in the effective action as $\sim Z(\phi) \partial_\mu \phi\partial^\mu \phi$ and evaluates $Z$ at a symmetry breaking stationary point of the effective action. This procedure is essential, e.g., in two-dimensional systems that undergo topological phase transitions, as the wave function renormalization is known to be diverging in the low temperature phase, which cannot be described in terms of perturbation theory \cite{gersdorff01}. Similarly,  in four dimensions, it has recently been shown that in the three flavor linear sigma model the coefficient of `t Hooft's determinant term also receives substantial contributions when evaluated at nonzero field \cite{fejos18}.

A similar treatment should be in order for Yukawa interactions, whose renormalization group flow vanishes at the one loop level, if complex scalar fields are coupled to the fermions and the field strength renormalization is trivial \cite{peskin}. In phenomenological investigations of the two and three flavor quark--meson models several papers have dealt with the nonperturbative renormalization of the Yukawa coupling. The essence of the corresponding calculations is that one determines the RG flow of the fermion--fermion--meson proper vertex, defined as $\delta^3 \Gamma/\delta_L\bar{\psi}\delta_R \psi \delta \phi$ in a given background ($\Gamma$ being the effective action), and then associates it with the flow of the Yukawa coupling itself. This obviously works for one flavor models \cite{gies04,braun09}, and even for the two flavor case \cite{berges99,pawlowski14}, in particular for models restricted to the $\sigma-\pi$ subsector, i.e., with $O(4)$ symmetry in the meson part of the theory. The reason why the approach has legitimacy in the latter case is that it turns out that the only way fluctuations generate new Yukawa--like operators in the quantum effective action is that the original Yukawa term is multiplied by powers of the quadratic $O(4)$ invariant of the meson fields. That is to say, one is allowed to choose, e.g., a vacuum expectation value for which all pseudoscalar ($\pi)$ fields vanish, but the scalar ($s$) one is nonzero, and associate $\sim\!s^2$ with the aforementioned quadratic invariant leading to the appropriate determination of the field dependence of the Yukawa coupling.

The situation is not that simple in case of three flavors, i.e., chiral symmetry of $U_L(3)\times U_R(3)$. Since the effective action automatically respects linearly realized symmetries of the Lagrangian, field dependence of any coupling must mean all possible functional dependence on chirally invariant operators. Because there are more invariants compared to the two flavor case, if one chooses a specific background for the evaluation of the RG flow of the fermion--fermion--meson vertex, one may arrive at an ``effective'' Yukawa interaction, but it will then mix in that specific background the contribution of different chirally invariant operators. The reason is that for three flavors other operators than that of the Yukawa term (multiplied by powers of the quadratic invariant) can get generated, which are presumably not even considered neither in the Lagrangian nor in the ansatz of the effective action. The newly emerging operators in principle cannot be distinguished from the Yukawa interaction if the background is too simple, and as it will be shown, a naive calculation squeezes inappropriate terms into the Yukawa flow, which should be separated via the choice of a suitable background and projected out.

Investigations of the three flavor case including the effects of the $U_A(1)$ anomaly were started by Mitter and Schaefer. In a first paper, no RG evolution of the Yukawa coupling was taken into account \cite{mitter14}. This approximation was carried over in several directions, including investigations regarding magnetic \cite{kamikado15} and topological \cite{jiang16} susceptibilities, mass dependence of the chiral transition \cite{resch17}, or quark stars \cite{otto20}.

A strategy for extracting the RG flow of the field-dependent Yukawa coupling was presented by Pawlowski and Rennecke, explicitly realized for the two flavor case \cite{pawlowski14}. In this paper, interaction vertices arising from the expansion of the field-dependent Yukawa coupling were also analyzed and it was found that higher order mesonic interactions to two quarks are gradually less important. These results were further used in recent ambitious computations of the meson spectra from effective models directly matched to QCD \cite{cyrol18,alkofer19}. The method of \cite{pawlowski14} was then applied for extracting the flow of the Yukawa coupling in the 2+1 flavor (realistic) quark--meson model \cite{rennecke17}. The prescription used in the latter calculation relies on a symmetry breaking background, from which a field independent Yukawa coupling is extracted. A similar approach has already been introduced in the seminal paper of Jungnickel and Wetterich \cite{jungnickel96}, where a diagonal symmetry breaking background was employed in calculating the RG equation of the Yukawa coupling (for general number of flavors).
As a generalization, here we offer a renormalization procedure of the field-dependent Yukawa coupling realizable in any symmetry breaking background through a chirally invariant set of operators.

More specifically, the aim of this paper is to show how to separate the field-dependent Yukawa interaction from those new fermion--fermion--multimeson couplings that unavoidably arise as three flavor chiral symmetry allows for a much larger set of operators in the infrared than considering only a field-dependent Yukawa coupling. For this, the right-hand side (rhs) of the evolution equation of the effective action will be evaluated in such a background that allows for expressing it in terms of explicitly invariant operators, as it was demonstrated for purely mesonic theories some time ago \cite{patkos12,grahl13,grahl14,fejos14}. We also note that apart from the above applications, recent studies on a lower Higgs mass bound \cite{gies17} and on quantum criticality \cite{vacca15} also deal with field-dependent Yukawa interactions.

The paper is organized as follows. In Sec. II, we present the model emphasizing its symmetry properties. We explicitly show what kind of new operators can be generated and set an ansatz for the scale-dependent effective action accordingly. Section III contains the explicit calculation of the flows generated by the field-dependent Yukawa coupling, and Sec. IV is devoted to an extended scenario where at the lowest order the newly established operators are also taken into account. In Sec. V, we present numerical evidence for the relevance of the proposed procedure, while the reader finds the summary in Sec. VI.

\section{Model and symmetry properties}

We are working with the $U_A(1)$ anomaly free three flavor quark--meson model, which is defined through the following Euclidean Lagrangian:
\bea
\label{Eq:Lag}
{\cal L}&=&\Tr(\partial_i M^\dagger \partial_i M) + m^2 \Tr (M^\dagger M) \nonumber\\
&+& \bar{\lambda}_1 \left(\Tr(M^\dagger M)\right)^2 +\bar{\lambda}_2\Tr (M^\dagger M M^\dagger M)\nonumber\\
&+&\bar{q}(\slashed{\partial}+g{M}_5)q,
\eea
where $M$ stands for the meson fields, $M=(s_a+i\pi_a)T_a$ ($T_a=\lambda_a/2$ are generators of $U(3)$ with $\lambda_a$ being the Gell-Mann matrices, $a=0...8$), $q^T=(u \hspace{0.08cm} d \hspace{0.08cm} s)$ are the quarks, and $M_5=(s_a+i\pi_a\gamma_5)T_a$. As usual, $m^2$ is the mass parameter and $\bar{\lambda}_1$, $\bar{\lambda}_2$ refer to independent quartic couplings. The fermion part contains $\slashed{\partial}\equiv \partial_i \gamma_i$, $\gamma_5=i\gamma_0\gamma_1\gamma_2\gamma_3$ with $\{\gamma_i\}$ being the Dirac matrices and the Yukawa coupling is denoted by $g$.

Concerning the meson fields, $U_L(3)\times U_R(3) \simeq U_V(3)\times U_A(3)$ chiral symmetry manifests itself as
\bea
M \rightarrow VMV^\dagger, \hspace{1cm} M \rightarrow A^\dagger M A^\dagger
\eea
for vector and axialvector transformations, respectively, $V=\exp(i\theta_V^a T_a)$, $A=\exp(i\theta_A^aT_a)$. This also implies that
\bea
M_5 \rightarrow V M_5 V^\dagger, \hspace{1cm} M_5 \rightarrow A_5^\dagger M_5 A_5^\dagger,
\eea
where $A_5=\exp(i\theta_A^aT_a\gamma_5)$.
As for the quarks, the transformation properties are
\bea
q \rightarrow Vq, \hspace{1cm} q \rightarrow A_5 q.
\eea
Note that, since $\{\gamma_i,\gamma_5\}=0$,
\bea
\bar{q} \rightarrow \bar{q} V^\dagger, \hspace{1cm} \bar{q} \rightarrow \bar{q}A_5.
\eea
These transformation properties guarantee that all terms in (\ref{Eq:Lag}) are invariant under vector and axialvector transformations.

The quantum effective action, $\Gamma$, built upon the theory defined in (\ref{Eq:Lag}) also has to respect chiral symmetry. That is to say, only chirally invariant combinations of the fields can emerge in $\Gamma$. For three flavors, there exist three independent invariants made out of the $M$ fields,
\bea
\rho&:=& \Tr (M^\dagger M), \nonumber\\
\tau&:=& \Tr (M^\dagger M - \Tr(M^\dagger M)/3)^2, \nonumber\\
\rho_3&:=& \Tr (M^\dagger M - \Tr(M^\dagger M)/3)^3,
\eea
where the last one is absent in (\ref{Eq:Lag}) due to perturbative UV renormalizability but nothing prevents its generation in the infrared. In principle, the chiral effective potential, $V$, defined via homogeneous field configurations, $\Gamma|_{\hom}=\int_x V$ has to be of the form $V=V(\rho,\tau,\rho_3)$. Obviously, the generalized Yukawa term,
\bea
\label{Eq:genyukawa}
g(\rho,\tau,\rho_3) \cdot\bar{q}M_5q
\eea
can (and does) also appear in $\Gamma$, but in this paper we restrict ourselves to
\bea
\label{Eq:Yuksimple}
g(\rho,\tau,\rho_3)\approx g(\rho).
\eea
This is motivated from a symmetry breaking point of view, i.e., without explicit symmetry breaking terms in ({\ref{Eq:Lag}), chiral symmetry breaks as $U_L(3)\times U_R(3) \rightarrow U_V(3)$ \cite{vafa84}, and for any background field respecting vector symmetries, $\tau\equiv 0 \equiv \rho_3$. We can think of $g(\rho)$ as
\bea
\label{Eq:gseries}
g(\rho)=\sum_{n=0}^{\infty} g_n \rho^n=(g_0+g_1\cdot \rho+...),
\eea
which shows that it actually resums operators of the form $\sim [\Tr(M^\dagger M)]^n \bar{q}M_5 q$. 

Note that, however, as announced in the Introduction, there are several other invariant combinations at a given order that are allowed by chiral symmetry and contribute to quark--meson interactions. Take the lowest nontrivial order, i.e., dimension 6. We have two invariant combinations:
\bea
\sim \!\rho \cdot\bar{q}M_5 q, \hspace{0.6cm} \sim\!\bar{q}M_5 M_5^\dagger M_5 q.
\eea
Obviously the first one is to be incorporated into the field-dependent Yukawa coupling, but not the second one. Note that one may think of the fifth Dirac matrix as the difference between left- and right-handed projectors, $\gamma_5=P_R-P_L$, and therefore, e.g., the first expression is  equivalent to $\bar{q}_L M q_R+\bar{q}_RM^\dagger q_L$ and the second to $\bar{q}_L M M^\dagger M q_R+\bar{q}_R M^\dagger M M^\dagger q_L$. Thus, one can trade the $M_5$ dependence to $M$ dependence in the invariants via working with left- and right-handed quarks separately.
Going to next-to-leading order, we see the emergence of the following new terms [note that for general configurations $\Tr(M_5^\dagger M_5)=\Tr (M^\dagger M)$]:
\bea
\sim \!\rho^2 \cdot \bar{q}M_5q, \hspace{0.15cm} \sim\!\rho \cdot \bar{q}M_5 M_5^\dagger M_5 q, \hspace{0.15cm}\sim\! \bar{q} M_5M_5^\dagger M_5 M_5^\dagger M_5 q, \nonumber\\
\eea
and we might even have $\sim\! \tau \cdot \bar{q}M_5q$, though we intend to drop such contributions, see approximation (\ref{Eq:Yuksimple}). If we are to extract the renormalization group flow of $g(\rho)$, then we need a strategy to distinguish each operator at a given order, in order to be able to resum only the Yukawa interactions, and not  the aforementioned new quark--quark--multimeson operators. Obviously, this procedure cannot be done through a one component background field, e.g., the vectorlike condensate $ M \sim s_0\cdot\mathbb{1}$, as it mixes up the corresponding operators, and one loses the chance to recombine the field dependence into actual invariant tensors.

The flow of the effective action is described by the Wetterich equation \cite{wetterich93,morris94}
\bea
\label{Eq:wet}
\partial_k \Gamma_k = \frac12 \int \Tr [(\Gamma_k^{(2)}+{\cal R}_k)^{-1}\partial_k {\cal R}_k],
\eea
where $\Gamma_k^{(2)}$ is the second functional derivative matrix of $\Gamma_k$ and ${\cal R}_k$ is the regulator function. One typically chooses it to be diagonal in momentum space, and set ${\cal R}_k$ as ${\cal R}_k(p,q)=(2\pi)^4 R_k(q)\delta (q+p)$. Evaluating (\ref{Eq:wet}) in Fourier space leads to
\bea
\label{Eq:wet2}
\partial_k \Gamma_k = \frac12 \int_p \Tr [(\Gamma_k^{(2)}+{\cal R}_k)^{-1}(p,-p)\partial_k R_k(p)].
\eea
In this paper, the equation will be evaluated only in spacetime-independent background fields; therefore, one assumes that $\Gamma_k^{(2)}$ is also diagonal in momentum space, $\Gamma_k^{(2)}(p,q)=(2\pi)^4\Gamma_k^{(2)}(q)\delta(q+p)$. In such backgrounds, it is sufficient to work with the effective potential, $V_k$, via $\Gamma_k|_{\hom}=\int_x V_k$, and (\ref{Eq:wet2}) leads to
\bea
\label{Eq:wet3}
\partial_k V_k &=& \frac12 \int_p \Tr \{(\Gamma_k^{(2)}(p_R))^{-1}\partial_k R_k(p)\},
\eea
where we have assumed that the regulator matrix $R_k$ is meant to replace everywhere $p$ with $p_R$, where $p_R$ is the regulated momentum through the $R_k$ function, i.e.,  $p_R^2=p^2+R_k(p)$, $p_{Ri}=p_R\hat{p}_i$.

The ansatz we choose for $\Gamma_k$ is called the local potential approximation, which consists of the usual kinetic terms and a local potential (in this case, specifically, a mesonic potential plus the Yukawa term),
\bea
\label{Eq:ansatz}
\Gamma_k=\int_x &&\Big[ \Tr(\partial_i M^\dagger \partial_i M)+ \bar{q}\slashed{\partial}q \nonumber\\
&&+V_{\ch,k}(\rho,\tau,\rho_3)+g_k(\rho)\bar{q}{M}_5q\Big]. \\
\nonumber
\eea
Using the earlier notation, $V_k=V_{\ch,k}+g_k(\rho)\bar{q}M_5 q$. Note that the chiral potential will not be considered in its full generality, but rather as $V_{\ch,k}=U_k(\rho)+C_k(\rho)\tau$. Flows of $U_k$ and $C_k$ can be found in \cite{fejos14}. Note that this construction could be easily extended with `t Hooft's determinant term describing the $U_A(1)$ anomaly, $\sim (\det M + \det M^\dagger)$, and the corresponding coefficient function. Investigations in this direction are beyond the scope of this paper.

\section{Flow of the field-dependent Yukawa coupling}

In this section, we evaluate (\ref{Eq:wet2}) in homogeneous background fields to obtain the flow of $g_k(\rho)$, defined in (\ref{Eq:ansatz}). Calculating $\Gamma_k^{(2)}$ using the ansatz (\ref{Eq:ansatz}}), one gets the following hypermatrix in the $\{s^a$, $\pi^a$, $\bar{q}^T$, $q\}$ multicomponent space:
\bea
\begin{pmatrix}
p^2+m_{s,k}^2 & m_{s\pi,k}^2 & -(\vec{g}^{(s)}_k q)^T & \bar{q}\vec{g}^{(s)}_k \\
m_{\pi s,k}^2 & p^2+m_{\pi,k}^2 & -(\vec{g}^{(\pi)}_kq)^T & \bar{q}\vec{g}^{(\pi)}_k\\
\vec{g}^{(s)}_k q & \vec{g}^{(\pi)}_kq & 0 & i\slashed{p}+g_k M_5\\
-(\bar{q}\vec{g}^{(s)}_k)^T & -(\bar{q}\vec{g}_k^{(\pi)})^T & -i\slashed{p}-g_k M_5 & 0\\
\end{pmatrix},
\nonumber\\
\eea
where $(g^{(s)}_k)_a=g_k(\rho)T_a+g_k'(\rho)s_aM_5$, $(g^{(\pi)}_k)_a=ig_k(\rho)T_a\gamma_5 +g'_k(\rho) \pi_a M_5$, and the mass matrices are defined through the following relations:
\bea
(m_{s,k}^2)_{ij}=\partial^2 V_k/\partial s_i\partial s_j, &&\quad (m_{\pi,k}^2)_{ij}=\partial^2 V_k/\partial \pi_i\partial \pi_j,\nonumber\\ (m^2_{s\pi,k})_{ij}&=&\partial^2 V_k/\partial s_i\partial \pi_j.
\eea
For practical purposes, it is worth to separate $\Gamma_k^{(2)}$ into three parts \cite{aoki97,jakovac13}, $\Gamma_k^{(2)}=\Gamma_{k,B}^{(2)}+\Gamma_{k,F}^{(2)}+\Gamma_{k,\mix}^{(2)}$, where the respective terms are defined as
\begin{widetext}
\bea
\Gamma_{k,B}^{(2)}&=&
\begin{pmatrix}
p^2+m_{s,k}^2 & m_{s\pi,k}^2 & 0 & 0 \\
m_{\pi s,k}^2 & p^2+m_{\pi,k}^2 & 0 & 0\\
0 & 0 & 0 & 0\\
0 & 0 & 0 & 0\\
\end{pmatrix},
\qquad \Gamma_{k,F}^{(2)}=
\begin{pmatrix}
0 & 0 & 0 & 0 \\
0 & 0 & 0 & 0 \\
0 & 0 & 0 & i\slashed{p}+g_k M_5\\
0 & 0 & -i\slashed{p}-g_k M_5 & 0\\
\end{pmatrix},
\nonumber\\
\Gamma_{k,\mix}^{(2)}&\equiv&
\begin{pmatrix}
0 & \Gamma_{k,FB}^{(2)} & \\
\Gamma_{k,BF}^{(2)} & 0 \\
\end{pmatrix}
=
\begin{pmatrix}
0 & 0 & -(\vec{g}^{(s)}_k q)^T & \bar{q}\vec{g}^{(s)}_k \\
0 & 0 & -(\vec{g}^{(\pi)}_kq)^T & \bar{q}\vec{g}^{(\pi)}_k\\
\vec{g}^{(s)}_k q & \vec{g}^{(\pi)}_kq & 0 & 0\\
-(\bar{q}\vec{g}^{(s)}_k)^T &  -(\bar{q}\vec{g}_k^{(\pi)})^T  & 0 & 0\\
\end{pmatrix}. \\
\nonumber
\eea
\end{widetext}
Using these notations, and by introducing the differential operator $\tilde{\partial}_k$, which, by definition acts only on the regulator function, $R_k$, (\ref{Eq:wet3}) can be conveniently reformulated. 
\begin{figure}[t]
\includegraphics[bb = 380 450 155 565,scale=0.85,angle=0]{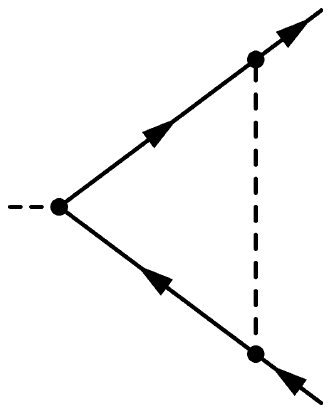}
\caption{One-loop triangle diagram that is responsible for flowing quark--meson interactions. The solid lines correspond to quarks, while the dashed ones represent mesons.}
\end{figure}  
\begin{figure}[t]
\includegraphics[bb = 140 460 145 525,scale=1.4,angle=0]{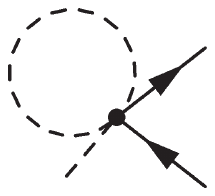}
\caption{One-loop tadpole diagram arising from the field-dependent Yukawa interaction, which allows the generation of new quark--meson vertices in the RG flow. The solid lines correspond to quarks, while the dashed ones represent mesons.}
\end{figure}  
We may use the matrix identity
\bea
&&\begin{pmatrix}
\Gamma_{k,B}^{(2)} & \Gamma_{k,BF}^{(2)} \\
\Gamma_{k,FB}^{(2)} & \Gamma_{k,F}^{(2)}
\end{pmatrix}
=\nonumber\\
&& \hspace{0.5cm}
\begin{pmatrix}
\Gamma^{(2)}_{k,B} & 0 \\
\Gamma^{(2)}_{k,FB} & 1
\end{pmatrix}
\begin{pmatrix}
1 & (\Gamma_{k,B}^{(2)})^{-1}\Gamma_{k,BF}^{(2)} \\
0 & \Gamma_{k,F}^{(2)}-\Gamma_{k,FB}^{(2)}(\Gamma_{k,B}^{(2)})^{-1}\Gamma_{k,BF}^{(2)}
\end{pmatrix}
\nonumber\\
\eea
and arrive at
\bea
\label{Eq:wet4}
\partial_k V_k &=& \frac12 \int_p \tilde{\partial}_k \Tr \log \Gamma_k^{(2)}(p_R)\nonumber\\
&&\hspace{-0.65cm}=\frac12 \int_p \tilde{\partial}_k \Tr \log \Gamma_{k,B}^{(2)}(p_R)+\frac12 \int_p \tilde{\partial}_k \Tr \log \Gamma_{k,F}^{(2)}(p_R)\nonumber\\
&&\hspace{-0.65cm}+\frac12 \int_p \tilde{\partial}_k \Tr \log [1- G_{k,F}(p_R)\Gamma_{k,FB}^{(2)}G_{k,B}(p_R)\Gamma^{(2)}_{k,BF}],\nonumber\\
\eea
where $G_{k,B}=(\Gamma_{k,B}^{(2)})^{-1}$, $G_{k,F}=(\Gamma_{k,F}^{(2)})^{-1}$ and the negative sign of the pure fermionic term is understood. We are interested in the flow of the Yukawa coupling; thus, we need to identify in the rhs of (\ref{Eq:wet4}) the operator $\sim \bar{q}M_5q$. This is in part generated from the leading term of the last contribution of the rhs of (\ref{Eq:wet4}),
\bea
\label{Eq:triangle}
-\frac{1}{2}\int_p \tilde\partial_k\textrm{Tr}[G_{k,F}(p_R)\Gamma^{(2)}_{k,FB}G_{k,B}(p_R)\Gamma^{(2)}_{k,BF}].
\eea
More specifically, it reads
\bea
\label{Eq:rhs}
&&-\bar{q}\int_p \tilde{\partial}_k \Big\{ (g^{(s)}_{k})_aS(p_R) G_{k,s}^{ab}(p_R)(g^{(s)}_{k})_b\nonumber\\
&&\hspace{1.3cm}+(g^{(\pi)}_{k})_aS(p_R) G_{k,\pi}^{ab}(p_R)(g^{(\pi)}_{k})_b\nonumber\\
&&\hspace{1.3cm}+(g^{(s)}_{k})_aS(p_R) G_{k,s\pi}^{ab}(p_R)(g^{(\pi)}_{k})_b\nonumber\\
&&\hspace{1.3cm}+(g^{(\pi)}_{k})_aS(p_R) G_{k,\pi s}^{ab}(p_R)(g^{(s)}_{k})_b\Big\} q,
\eea
where $S_k(p)=(i\slashed{p}+g_kM_5)^{-1}$ is the fermion propagator without doubling, while $G_{k,s}$, $G_{k,\pi}$, $G_{k,s\pi}$ and $G_{k,\pi s}$ are the respective $9\times 9$ submatrices of $G_{k,B}$ in a purely bosonic background (see Appendix B). Note that when expanding $S$ around zero field, at the leading order (\ref{Eq:rhs}) leads diagrammatically to the commonly known term seen in Fig. 1.

In principle, the evaluation of (\ref{Eq:rhs}) goes as follows. First, one evaluates $\Gamma_{k,B}$, then inverts it to obtain $G_{k,s}$, $G_{k,\pi}$, $G_{k,s\pi}$ and $G_{k,\pi s}$. Second, one inverts the fermion propagator, and finally performs all matrix multiplications to get the operator that is sandwiched by $\bar{q}$ and $q$. Note that the procedure turns out to be too complicated to be carried out in a general background, but as we explain in the next subsections, fortunately we will not need to evaluate (\ref{Eq:rhs}) in its full generality. 

On top of the above, via a field-dependent Yukawa coupling there is also contribution from the first term of (\ref{Eq:wet4}), as meson masses get modified by a fermionic background (see Appendix B) leading to the possibility of generating a $\sim \bar{q}M_5 q$ term in the effective action; see Fig. 2. Generically, this contributes to the rhs of (\ref{Eq:wet4}) as
\begin{widetext}
\bea
\label{Eq:rhs2}
\includegraphics[bb = 130 480 175 525,scale=0.45,angle=0]{fig2.pdf}=\bar{q} \frac{g_k'(\rho)}{2}\int_p \tilde{\partial}_k \Big\{ &&G_{k,s}^{ab}(p_R)\Big(\frac{\partial \rho}{\partial s_a}\frac{\partial M_5}{\partial s_b}+\frac{\partial \rho}{\partial s_b}\frac{\partial M_5}{\partial s_a}+\delta_{ab}M_5\Big) +G_{k,\pi}^{ab}(p_R)\Big(\frac{\partial \rho}{\partial \pi_a}\frac{\partial M_5}{\partial \pi_b}+\frac{\partial \rho}{\partial \pi_b}\frac{\partial M_5}{\partial \pi_a}+\delta_{ab}M_5\Big), \nonumber\\
&&\!\!\!\!\!+G_{k,s\pi}^{ab}(p_R)\Big(\frac{\partial \rho}{\partial \pi_b}\frac{\partial M_5}{\partial s_a}+\frac{\partial \rho}{\partial s_a}\frac{\partial M_5}{\partial \pi_b}\Big)+G_{k,\pi s}^{ab}(p_R)\Big(\frac{\partial \rho}{\partial \pi_a}\frac{\partial M_5}{\partial s_b}+\frac{\partial \rho}{\partial s_b}\frac{\partial M_5}{\partial \pi_a}\Big)\Big\} q \nonumber\\
&&\hspace{-2.4cm}+\bar{q} \frac{g_k''(\rho)}{2} \int_p \tilde{\partial}_k \Big\{ G_{k,s}^{ab}(p_R) \frac{\partial \rho}{\partial s_a}\frac{\partial \rho}{\partial s_b}+G_{k,\pi}^{ab}(p_R) \frac{\partial \rho}{\partial \pi_a}\frac{\partial \rho}{\partial \pi_b}+G_{k,s\pi}^{ab}(p_R) \frac{\partial \rho}{\partial s_a}\frac{\partial \rho}{\partial \pi_b}+G_{k,\pi s}^{ab}(p_R) \frac{\partial \rho}{\partial s_b}\frac{\partial \rho}{\partial \pi_a}\Big\}M_5 q. 
\eea
\end{widetext}

\subsection{Flows in the symmetric phase}

As a first step, we are interested in the RG flows in the symmetric phase, i.e., where they are obtained at zero field. Note that, however, this still necessitates the evaluation of the rhs of the flow equation (\ref{Eq:rhs}) in a nonzero background, but if we are interested in the flowing couplings of dimension 6 operators, it is sufficient to perform all computations at the cubic order in the meson fields.
That is to say, when all propagators are expanded around $M =0$ (see Appendix B for the general formulas), one is allowed to work with them at the following accuracy:
\begin{subequations}
\label{Eq:propagator-expansion}
\bea
G^{ab}_{k,s}(p)&=&\frac{1}{p^2+U_k^\prime}\delta_{ab}-\frac{1}{(p^2+U_k^\prime)^2} \nonumber\\
&\times&\Big(s_as_b U_k'' + \frac{\partial^2\tau}{\partial s_a\partial s_b}C_k\Big),\\
G^{ab}_{k,\pi}(p)&=&\frac{1}{p^2+U_k^\prime}\delta_{ab}-\frac{1}{(p^2+U_k^\prime)^2} \nonumber\\
&\times&\Big(\pi_a \pi_b U_k'' + \frac{\partial^2\tau}{\partial \pi_a\partial \pi_b}C_k\Big),\\
G^{ab}_{k,s\pi}(p)&=&-\frac{1}{(p^2+U_k^\prime)^2}\Big(s_a \pi_b U_k'' + \frac{\partial^2\tau}{\partial s_a\partial \pi_b}C_k\Big),\nonumber\\
\\
S_k(p)&=& -\frac{i\slashed{p}}{p^2}+\frac{g_k M_5^{\dagger}}{p^2},
\eea
\end{subequations}
where the coefficient functions (i.e., $U_k', U_k'', C_k$) are evaluated at zero field. Note that in this subsection all computations are performed without specifying any background field and keeping its most general form. Terms that contain higher derivatives of $C_k$ are left out since their coefficients would lead to subleading (higher than cubic power) contributions in the background fields in (\ref{Eq:rhs}) and (\ref{Eq:rhs2}). 

Note that the scalar and pseudoscalar meson propagators do not mix in the symmetric phase and are degenerate; therefore, in the ${\cal O}(g_k^3$) term there is a relative $(-1)$ factor between their contributions in (\ref{Eq:rhs}), which exactly cancels the term linear in the mesonic fields. In the piece of the integrand that is quadratic in them, the quark momentum is odd; therefore, upon integration, it also vanishes. Together with the ${\cal O}(g_k^2g_k')$ piece, the leading contribution of (\ref{Eq:rhs}) is determined by
\bea
&&\hspace{-0.4cm}\includegraphics[bb = 160 500 305 565,scale=0.25,angle=0]{fig1.pdf}=\bar{q} \int_p \tilde{\partial}_k \frac{1}{p_R^2(p_R^2+U_k')^2} \Big\{g_k^3 T_a M_5^\dagger \nonumber\\
&&\times \Big[(s_as_b - \pi_a\pi_b)U_k''+\frac{\partial^2 \tau}{\partial s_a \partial s_b}C_k - \frac{\partial^2 \tau}{\partial \pi_a \partial \pi_b}C_k \nonumber\\
&&+i\gamma_5 (s_a\pi_b+s_b\pi_a)U_k''+i\gamma_5 C_k \Big(\frac{\partial^2 \tau}{\partial s_a \partial \pi_b}+\frac{\partial^2 \tau}{\partial \pi_a \partial s_b}\Big)\Big]T_b\Big\}q\nonumber\\
&&+\bar{q} \int_p \tilde{\partial}_k \frac{-1}{p_R^2(p_R^2+U_k')} \Big\{2g_k^2g'_k (s_a+i\pi_a\gamma_5)T_a M_5^\dagger \nonumber\\
&&\hspace{4.8cm}\times(s_b+i\pi_b\gamma_5)T_b\Big\}q.
\eea
Exploiting various identities of the $U(3)$ algebra (see Appendix A), one can perform the algebraic evaluation of the integrands to arrive at
\bea
\label{Eq:rhs1res}
&&\hspace{-1.8cm}\includegraphics[bb = 160 500 305 565,scale=0.25,angle=0]{fig1.pdf}= \int_p\tilde\partial_k\frac{1}{p_R^2(p_R^2+U_k^\prime)^2}\nonumber\\
&\times& \Big[\big(g_k^3 U_k^{\prime\prime}-\frac{2}{3}g_k^3C_k\big)\bar{q} \big(M_5M_5^\dagger-\frac{\rho}{3}\big) M_5q \nonumber\\
&+&\hspace{0.1cm}\big(\frac13g_k^3U_k''+\frac{16}{9}g_k^3C_k'\big)\rho \bar{q} M_5 q\Big]\nonumber\\
&&\hspace{-1cm}+\int_p\tilde\partial_k\frac{1}{p_R^2(p_R^2+U_k^\prime)}\Big[(-2g_k^2g_k')\bar{q} \big(M_5M_5^\dagger-\frac{\rho}{3}\big) M_5q\nonumber\\
&&\hspace{2.5cm}-\frac23g_k^2g_k\bar{q}\rho M_5 q\Big].
\eea
Eq. (\ref{Eq:rhs1res}) clearly shows the appearance of a new dimension 6 operator ($\sim \! \bar{q}M_5 M_5^\dagger M_5 q$), which was absent in the ansatz of the effective action. Before discussing this result, let us turn to the contribution of (\ref{Eq:rhs2}). Using, again, expressions (\ref{Eq:propagator-expansion}), a straightforward calculation leads to
\bea
\label{Eq:rhs2res}
&&\includegraphics[bb = 130 480 175 525,scale=0.45,angle=0]{fig2.pdf}=\int_p \tilde{\partial}_k \frac{1}{p_R^2+U_k'} \Big[10g_k' \bar{q}M_5q+ g_k'' \rho \bar{q}M_5 q\Big]\nonumber\\
&&\hspace{-0.4cm} +\int_p \tilde{\partial}_k \frac{1}{(p^2_R+U_k')^2}\Big[(-\frac{16}{3}g_k'C_k-3g_k'U_k'')\rho \bar{q}M_5q\nonumber\\
&&\hspace{2.5cm} -6g_k'C_k \bar{q}\big(M_5M_5^\dagger-\rho/3\big) M_5 q\Big],
\eea
where the coefficient functions (i.e., $U_k', g_k, g_k', g_k'', C_k$) are, once again, evaluated at zero field. The flow of the effective potential is the sum of (\ref{Eq:rhs1res}) and (\ref{Eq:rhs2res}).

At this point, it is clear that the flow equation does not close in the sense that the ansatz (\ref{Eq:ansatz}) does not contain the operator $\sim \!\bar{q}M_5 M_5^\dagger M_5 q$; therefore, for consistency reasons, it has to be dropped also in the rhs of (\ref{Eq:wet4}). Instead of doing so, one may choose to project out the $\sim \! \bar{q}( M_5 M_5^\dagger - \rho/3) M_5 q$ term, because in the symmetry breaking pattern $U_L(3)\times U_R(3) \rightarrow U_V(3)$ that is the actual combination, which vanishes. This choice should lead to the appropriate definition of the field-dependent flowing Yukawa coupling if no explicit symmetry breaking terms are present. Note, however, that, once finite quark masses and the corresponding explicit breaking terms are introduced, in the minimum point of the effective potential $M_5 M_5^\dagger \neq \rho/3$, and the aforementioned projection can be regarded as somewhat arbitrary. A more appropriate treatment is to include $\sim\!\bar{q}M_5 M_5^\dagger M_5 q$ in the ansatz of the effective action in the first place (see Sec. IV). 

We close this subsection by listing the coupled flow equations for the Yukawa coupling and its derivative evaluated at zero field, i.e., in the symmetric phase, when  $\sim \! \bar{q}( M_5 M_5^\dagger - \rho/3) M_5 q$ is projected out. Dropping $g_k''$ in order to close the system of equations, and by using the definitions of (\ref{Eq:gseries}), (\ref{Eq:rhs1res}), and (\ref{Eq:rhs2res}), we are led to
\begin{subequations}
\label{Eq:g0g1flows}
\bea
\label{Eq:g0flow}
\partial_k g_{0,k}&=&10g_{1,k} \int_p \tilde{\partial}_k \frac{1}{p_R^2+U_k'}, \\
\label{Eq:g1flow}
\partial_k g_{1,k}&=&\Big(\frac13 g_{0,k}^3 U_k''+\frac{16}{9}g_{0,k}^3C_k\Big)\int_p \tilde{\partial}_k \frac{1}{p_R^2(p_R^2+U_k')^2}\nonumber\\
&-&\Big(\frac{16}{3}g_{1,k}C_k+3g_{1,k}U_k''\Big)\int_p\tilde{\partial}_k \frac{1}{(p^2_R+U_k')^2} \nonumber\\
&-&\frac23 g_{0,k}^2g_{1,k}\int_p \tilde{\partial}_k \frac{1}{p_R^2(p_R^2+U_k')}.
\eea
\end{subequations}

\subsection{Flows in the broken phase}

Now, we explore how the flowing couplings depend on the fields, more specifically, as described in (\ref{Eq:Yuksimple}), their $\rho$ dependence will be determined.

At this point, we specify a background in which we will evaluate the flow equation, as in case of general $\{M$, $\bar{q}$, $q\}$ configurations, there is no hope that the broken symmetry phase calculations can be done explicitly. The nice thing, however, is that it is sufficient to work in a restricted background. We, again, definitely need to assume nonzero $\bar{q}$, $q$, and then we specify the $M=s_0T_0+s_8T_8$ (physical) condensate. This is one of the minimal choices, which allows for a unique restoration of the $\sim \bar{q}M_5 M_5^\dagger M_5 q$ and $\sim \bar{q} M_5 q$ operators in the rhs of the flow equation (a one component condensate would definitely not allow to do so). We have checked explicitly with other backgrounds the uniqueness of the results, and as expected, found agreement.

As outlined above, the first step is to evaluate the two--point functions $\Gamma_{k,s}^{(2)}$, $\Gamma_{k,\pi}^{(2)}$ (note that in the current background there is no $s-\pi$ mixing), see details in Appendix B, in particular (\ref{Eq:gammas}) and (\ref{Eq:gammas2}). Then one inverts these matrices to obtain $G_{k,s}$ and $G_{k,\pi}$. They are diagonal except for the $0$--$8$ sectors. Similarly, for the fermion propagator, one has $S^{-1}_k(p)=(i\slashed{p}+g_k(\rho) M_5)$, which leads to the inverse
\bea
S_k(p)=\frac{-i\slashed{p}+g_k(\rho)M_5^\dagger}{p^2+g_k(\rho) M_5 M_5^\dagger}.
\eea
Since in the background in question $[S_k,M_5]=0$, Eq. (\ref{Eq:rhs}) becomes
\bea
\label{Eq:rhsspec}
&&-\bar{q} \int_p \tilde{\partial}_k \Big\{g_k^2(\rho)S_k(p_R)T_a T_b\big(G_{k,s}^{ab}(p_R)-G_{k,\pi}^{ab}(p_R)\big) \nonumber\\
&&\hspace{1.6cm}+g_k(\rho)g_k'(\rho)S_k(p_R)M_5(s_aT_b+s_bT_a) G_{k,s}^{ab}(p_R) \nonumber\\
&&\hspace{1.6cm}+(g_k'(\rho))^2S_k(p_R)M_5M_5s_as_b G_{k,s}^{ab}(p_R) \Big\} q. 
\eea
Similarly, exploiting the choice of the simplified background, (\ref{Eq:rhs2}) becomes
\bea
\label{Eq:rhs2spec}
&&\bar{q}\frac{g_k''(\rho)}{2}\int_p \tilde{\partial}_k G_{k,s}^{ab}(p_R)s_a s_b q\nonumber\\
&+&\bar{q} \frac{g_k'(\rho)}{2}\int_p \tilde{\partial}_k \Big(G_{k,s}^{ab}(p_R)(s_aT_b+s_bT_a)\nonumber\\
&&\hspace{1.6cm}+(G_{k,s}^{aa}(p_R)+G_{k,\pi}^{aa}(p_R))M_5\Big)q.
\eea
It is important to mention that one does not need to work with general $s_0$, $s_8$ background values, but can safely assume that $s_8 \ll s_0$, and thus expand both (\ref{Eq:rhsspec}) and (\ref{Eq:rhs2spec}) in terms of $s_8$ and work in the leading order. This simplification still uniquely allows for identifying the flow of the $\sim\!\bar{q} M_5 q$ and $\sim \!\bar{q} (M_5 M_5^\dagger -\rho/3) M_5 q$ operators. The latter, as expected, pops up again [it is inherently of ${\cal O}(s_8)$]; thus, a two component background is indeed necessary to obtain each flow. The sum of (\ref{Eq:rhsspec}) and (\ref{Eq:rhs2spec}) leads to
\begin{widetext}
\bea
&&\int_p \tilde{\partial}_k \Bigg[ \frac{3g_k^3/2}{(p_R^2+g_k^2\rho/3)(p_R^2+U_k')}-\frac{4g_k^3/3}{(p_R^2+g_k^2\rho/3)(p_R^2+U_k'+4C_k\rho/3)}-\frac{g_k^3/6+2\rho g_k^2g_k'/3+2\rho^2g_k g_k'^2/3}{(p_R^2+g_k^2\rho/3)(p_R^2+U_k'+2\rho U_k'')} \nonumber\\
&&\hspace{1cm} +\frac{9g_k'/2}{p_R^2+U_k'}+\frac{4g_k'}{p_R^2+U_k'+4C_k\rho/3}+\frac{3g_k'/2+\rho g_k''}{p_R^2+U_k'+2\rho U_k''}\Bigg]\bar{q}M_5 q\nonumber
\eea
\bea
\label{Eq:npflow}
&&+\int_p \tilde{\partial}_k \Bigg[ -\frac{2g_k'(3C_k+2C_k'\rho)}{(p_R^2+U_k'+4C_k\rho/3)(p_R^2+U_k'+2\rho U_k'')}-\frac{8g_kC_k^3g_k^2\rho^2/3}{(p_R^2+g_k^2\rho/3)(p_R^2+U_k')^2(p_R^2+4C_k\rho/3+U_k')^2} \nonumber\\
&&\hspace{1.4cm}-\frac{16g_k^3U_k''C_k^2 \rho^2}{(p_R^2+g_k^2\rho/3)(p_R^2+U_k')(p_R^2+U_k'+4C_k\rho/3)^2(p_R^2+U_k'+2\rho U_k'')}-\frac{2g_k^2g_k'(p_R^2+U_k')}{(p_R^2+g_k^2\rho/3)(p_R^2+U_k'+4C_k\rho/3)^2} \nonumber\\
&&\hspace{1.4cm}+\frac{g_k^3U_k''(p_R^2+U_k')+\big(-\frac{56}{9}g_k^3C^2_k+\frac{g_k}{3} (p_R^2+U_k')(4C_k'g_k^2-6g_k'^2(p_R^2+U_k')+12g_kg_k'U_k'') \big)\rho}{(p_R^2+g_k^2\rho/3)(p_R^2+U_k'+4C_k\rho/3)^2(p_R^2+U_k'+2\rho U_k'')} \nonumber\\
&&\hspace{1.4cm}+\frac{\big(\frac{16}{9}g_k^2g_k'C^2_k+\frac83 g_k^2g_k'C_k'(p_R^2+U_k')+\frac{16}{9}C_kg_k(C_k'g_k^2-3g_k'^2(p_R^2+U_k))\big)\rho^2+\frac{32}{9}g_kg_k'C_k(C_k'g_k-C_kg_k')\rho^3}{(p_R^2+g_k^2\rho/3)(p_R^2+U_k'+4C_k\rho/3)^2(p_R^2+U_k'+2\rho U_k'')}  \nonumber\\
&&\hspace{1.4cm}+\frac{\frac23g_k^3U_k'C_k-\frac23g_k^3C_k(p_R^2+U_k')-\frac23(2C_kg_k^2g_k'(p_R^2+U_k)-g_k^4g_k'(p_R^2+U_k')-2C_kg_k^2g_k'U_k'+2C_kg_k^3U_k'')\rho}{(p_R^2+g_k^2\rho/3)^2(p_R^2+U_k'+4C_k\rho/3)^2}\nonumber\\
&&\hspace{1.4cm}+\frac{\frac83 g_k^2g_k'C_kU_k''\rho^2+\frac{16}{9}C_kg_k^3g_k'^2\rho^3}{(p_R^2+g_k^2\rho/3)^2(p_R^2+U_k'+4C_k\rho/3)^2}+\frac{\big(\frac43 C_kg^3_kU_k'U_k''-\frac13g_k^5 U_k''(p_R^2+U_k')\big)\rho+A_2\rho^2+A_3\rho^3+A_4\rho^4}{(p_R^2+g_k^2\rho/3)^2(p_R^2+U_k'+4C_k\rho/3)^2(p_R^2+U_k'+2\rho U_k'')}\Bigg]\nonumber\\
&&\hspace{1.4cm}\times \bar{q}\big(M_5M_5^\dagger-\rho/3\big) M_5 q,
\eea
where
\bea
A_2&=&\frac{8}{27}g_k^5C_k+\frac23 g_k^3(p_R^2+U_k')(g_k'^2(p_R^2+U_k')-2g_kg_k'U_k'')+\frac43 g_k^4g_k'C_k(p_R^2+U_k')\nonumber\\
&-&\frac43 g_k^5C_kU_k'' -\frac83 g_k^2g_k'U_k'U_k''C_k+\frac83 g_k^3C_kU_k''^2, \nonumber\\
A_3&=&\frac{32}{27}g_k^4g_k'C_k^2-\frac{16}{3}g_k^2g_k'C_kU_k''^2,\qquad A_4=\frac{32}{27}g_k^3g_k'^2C_k(C_k-3U_k'').\nonumber
\eea
\end{widetext}
As in the previous subsection, we may drop the second term in (\ref{Eq:npflow}) due to consistency and thus the first term provides the result for the fully nonperturbative flow of the Yukawa coupling. One also notes that if the term in question is not projected out and considered in e.g. a $(s_0,s_8)$ (physical) background, then it cannot even be interpreted as a standard Yukawa term, as in such backgrounds $(M_5M_5^\dagger-\rho/3) \sim T_8$. The presence of such contribution would force us to introduce in addition a different Yukawa coupling and treat it similarly to as it was an explicit symmetry breaking term.

The expression obtained for the flow of the Yukawa coupling $g_k(\rho)$ is thought to be a resummation of operators as described below (\ref{Eq:gseries}). As such, one is able to determine how strong the interaction is between quarks and mesons as, e.g., the chiral condensate evaporates at high temperature. In principle, $g_k(\rho)$ also depends explicitly on the temperature, but one expects that the decrease in $\rho$ is more important \cite{fejos18}. For phenomenology, one needs to evaluate $g_k(\rho)$ and its derivatives at the minimum point of the effective potential, $\rho=\rho_{0,k}$. In this case, one can also consider the broken phase expansion
\bea
\label{Eq:gexpan}
g_k(\rho)=g_{0,k}+g_{1,k}\cdot(\rho-\rho_{0,k})+...,
\eea
which, via (\ref{Eq:npflow}) defines the broken phase flows of the $g_{0,k}$ and $g_{1,k}$ couplings. They can be found in Appendix C.

\section{Effect of the $M_5 M_5^\dagger M_5$ operator}

Motivated by the nonuniqueness of the definition of the field-dependent Yukawa coupling in the earlier setting, now we investigate the case when the term $\int_x g_{53,k} \bar{q} M_5 M_5^\dagger M_5 q$ is added to the ansatz (\ref{Eq:ansatz}) of the effective action (without subtracting $\rho\bar qM_5q/3$ from it). Here $g_{53,k}$ is a new running coupling constant, and we will not be considering its field dependence. Also, we restrict ourselves to calculations in the symmetric phase; therefore, a similar treatment as of in Sec. IIIA is in order. All calculations presented there are still valid, but there is one more term contributing to rhs of the flow equation of the effective action, coming from the tadpole diagram of Fig. 2 [see the mass matrices in (\ref{Eq:deltamass53})],
\bea
&&\bar{q} \frac{g_{53}}{2}\!\!\int_p\!\! \tilde{\partial}_k [G_{k,s}^{ij}(p_R)\big(\big\{M_5,\{T_i,T_j\}\big\}+T_iM_5^\dagger T_j + T_j M_5^\dagger T_i\big) \nonumber\\
&&\hspace{0.5cm}+G_{k,\pi}^{ij}(p_R)(\big\{M_5,\{T_i,T_j\}\big\}-T_iM_5^\dagger T_j - T_j M_5^\dagger T_i) \nonumber\\
&&\hspace{0.5cm}+G_{k,s\pi}^{ij}(p_R)i\gamma_5(\big[M_5,[T_i,T_j]\big]+T_i M_5^\dagger T_j + T_j M_5^\dagger T_i)\nonumber\\
&&\hspace{0.5cm}+G_{k,\pi s}^{ij}(p_R)i\gamma_5(\big[M_5,[T_j,T_i]\big]+T_i M_5^\dagger T_j + T_j M_5^\dagger T_i)]q.\nonumber\\
\eea
Substituting the propagators from Eqs. (\ref{Eq:propagator-expansion}), without specifying an actual background field, straightforward calculations lead to
\bea
\label{Eq:flowg53}
&&6g_{53,k}\bar{q}M_5q\int_p \tilde{\partial}_k\frac{1}{p_R^2+U_k'}-\big[ 2g_{53,k}C_k\rho \bar{q}M_5 q\nonumber\\
&&+(3g_{53,k}U_k''+10g_{53,k}C_k\big)\bar{q}M_5 M_5^\dagger M_5q\big]\int_p \tilde{\partial}_k \frac{1}{(p_R^2+U_k')^2}. \nonumber\\
\eea
Note that all couplings (i.e., $U_k'$, $U_k''$, $C_k$) are evaluated at zero field as we have worked in the symmetric phase. One observes that even though the original (field independent) Yukawa coupling does not flow in the symmetric phase, once one includes the $g_{53}$ coupling, it does run with respect to the scale. Using (\ref{Eq:g0flow}), (\ref{Eq:g1flow}), and (\ref{Eq:flowg53}), we arrive at the following system of equations for $g_{0,k}$, $g_{1,k}$, and $g_{53,k}$:
\bea
\partial_k g_{0,k}&=&(10g_{1,k}+6g_{53,k})\int_p \tilde{\partial}_k\frac{1}{p_R^2+U_k'}, \\
\partial_k g_{1,k}&=&-\Big(\frac{10}{3}g_{1,k}C_k+3g_{1,k}U_k''+2g_{53,k}C_k\Big)\nonumber\\
&\times&\!\!\!\int_p \tilde{\partial}_k\frac{1}{(p_R^2+U_k')^2}+ 2g_{0,k}^3C_k\int_p \tilde{\partial}_k \frac{1}{p_R^2(p_R^2+U_k')^2},\nonumber \\ \\
\partial_k g_{53,k}&=&-(6g_{1,k}C_k+10g_{53,k}C_k+3g_{53,k}U_k'') \nonumber\\
&\times&\int_p \tilde{\partial}_k \frac{1}{(p_R^2+U_k')^2}\nonumber\\
&+&(g_{0,k}^3U_k''-\frac23 g_{0,k}^3 C_k) \int_p \tilde{\partial}_k \frac{1}{p_R^2(p_R^2+U_k')^2} \nonumber\\
&-&2g^2_{0,k}g_{1,k}\int_p \tilde{\partial}_k \frac{1}{p_R^2(p_R^2+U_k')}. 
\eea
We wish to note that based on Sec. IIIB, a broken phase calculation can also be done. This, however, leads to such complicated formulas that we do not go into the details.

\section{Numerics}

Though a complete solution of the flow equation for the effective action is beyond the scope of the paper, in this section we wish to provide at least a convincing evidence of how important it is to consider the field-dependent version of the Yukawa interaction. We will be investigating the approximation scheme described in Sec. IIIC and solve the flow equation for the Yukawa coupling when expanded around the minimum point of the potential. That is, we are dealing with (\ref{Eq:brokeng0}) and (\ref{Eq:brokeng1}) numerically. In this section, we employ Litim's optimal regulator, $R_k(p)=(k^2-p^2)\Theta(k^2-p^2)$; thus, $p_R^2=k^2$, if $p<k$.

Let us imagine that we add the following symmetry breaking terms to the Lagrangian, ${\cal L}_h=-(h_0s_0+h_8s_8)$, where the $h_i (i=0,8)$ are external fields without scale dependence. Note that adding these terms do not change any of the RG flows. By introducing the standard nonstrange-strange basis as $h_{\ns}=h_0\sqrt{2/3}+h_8/\sqrt3$, $h_{\s}=h_0/\sqrt3-h_8\sqrt{2/3}$, the partially conserved axialvector current relations yield
\bea
m_{\pi}^2f_{\pi}=h_{\ns}, \quad m_K^2f_K=\frac{h_{\ns}}{2}+\frac{h_s}{\sqrt2},
\eea
where $m_{\pi}$ and $m_K$ are the pion and kaon masses, respectively. Their experimental values are $m_{\pi}\approx 140 \MeV$ and $m_K \approx 494 \MeV$, while the corresponding decay constants read $f_{\pi} \approx 93 \MeV$, $f_K \approx 113 \MeV$. These considerations lead to
\bea
\label{Eq:h}
h_{\ns}=m_{\pi}^2f_{\pi}, \quad h_{\s}=\frac{1}{\sqrt2}(2m_K^2f_K-m^2_{\pi}f_{\pi}),
\eea
i.e.,
\bea
h_0&=&\sqrt{\frac23}\big(m_{\pi}^2f_{\pi}/2+m_K^2f_K), \nonumber\\
h_8&=&\frac{2}{\sqrt3}\big(m_{\pi}^2f_{\pi}-m_K^2f_K).
\eea
Now we can make use of the chiral Ward identities
\begin{subequations}
\label{Eq:chWard}
\bea
\frac{\partial V_{\ch,k=0}}{\partial s_{\ns}}&=&m_{\pi}^2s_{\ns}-h_{\ns}, \\
\frac{\partial V_{\ch,k=0}}{\partial s_{\s}}&=&\frac{m_K^2-m_{\pi}^2}{\sqrt2}s_{\ns}+m_K^2s_{\s}-h_{\s},
\eea
\end{subequations}
where the $s_{\ns}$ and $s_{\s}$ condensates are defined analogously to their respective external fields. If we combine (\ref{Eq:chWard}) with (\ref{Eq:h}), we arrive at the conclusion that irrespectively of the remaining model parameters (in particular, on including the axial anomaly), in the minimum point of the effective action
\bea
s_{\ns,\min}=f_{\pi}, \quad s_{\s,\min}=\sqrt2(f_K-f_{\pi}/2).
\eea
That is to say, in the minimum point the $\rho$ invariant is $\rho_0\equiv \rho_{\min}=(s_{\ns,\min}^2+s_{\s,\min}^2)/2 \approx (93.5 \MeV)^2$.

Since we are interested in a rough estimate, (\ref{Eq:brokeng0}) and (\ref{Eq:brokeng1}) will be solved such that the scale dependence of the chiral potential is neglected, and we are only after $g_{0,k}$ (and $g_{1,k}$) at $k=0$. The expansion of the chiral potential around the minimum reads
\bea
V_{\ch}(\rho)=U'(\rho_0)(\rho-\rho_0)+U''(\rho_0)(\rho-\rho_0)^2/2+C(\rho_0)\tau. \nonumber\\
\eea
The pion and kaon masses in the minimum are
\begin{subequations}
\bea
m_{\pi}^2&=&U'+\frac{C}{6}(s_{\ns,\min}^2-2s_{\s,\min}^2),\\
m_{K}^2&=&U'+\frac{C}{6}(s_{\ns,\min}^2-3\sqrt2 s_{\ns,\min} s_{\s,\min}+4 s_{\s,\min}^2),\nonumber\\
\eea
\end{subequations}
which using physical masses lead to $U'\approx 0.147\GeV^2$ and $C \approx 84.37$. We still need $U''$, which is determined via the light scalar ($\sigma$) mass. Its expression reads
\bea
m_{\sigma}^2&=&U'+\frac12U''(s_{\ns\min}^2+s_{\s,\min}^2) \nonumber\\
&+&\frac{C}{12}(s_{\ns,\min}^2+10s_{\s,\min}^2)-\frac{1}{12}\sqrt{D},  \\
D&=&(5C+6U'')^2s_{\ns,\min}^4+4(7C+3U'')^2s_{\s,\min}^4\nonumber\\
&+&4(19C^2+105CU''-18U''^2)s_{\s,\min}^2s_{\ns,\min}^2.\nonumber
\eea
Setting $10\lesssim U'' \lesssim 20$ yields $469 \MeV \lesssim m_{\sigma} \lesssim 594 \MeV$, which is a suitable interval for the physical value of $m_\sigma$. 

For several different parameters, we show in Table I how the fluctuation corrected Yukawa coupling, $g_{0,k=0}$ differs from its UV value, $g_{0,k=\Lambda}$. The initial value of the $g_{1,k}$ was set to be zero at $k=\Lambda$. Note that without field dependence, $g_{0,k}$ would not flow with respect to the scale at all and stayed at its initial value. Therefore, comparing the initial value with the one obtained at $k=0$ shows the importance of considering a field-dependent Yukawa coupling.

\begin{table}[t]
\centering
\vspace{0.2cm}
  \begin{tabular}{ c | c | c | c}
    $U''$ & $g_{0,k=\Lambda}$ & $g_{0,k=0}$ & $\Delta g$ \\ \hline
     $10$ & 5 & 6.0 & 16\% \\ \hline
    $10$ & 10 & 14.4 & 31\% \\ \hline
    $10$ & 15 & 22.7 & 34\% \\ \hline
    $10$ & 20 & 30.3 & 34\% \\ \hline
  \end{tabular} \hspace{0.4cm}
  \begin{tabular}{c|c|c|c}
  $U''$ & $g_{0,k=\Lambda}$ & $g_{0,k=0}$ & $\Delta g$ \\ \hline
  $20$ & 5 & 6.0 & 16\% \\ \hline
    $20$ & 10 & 14.1 & 29\% \\ \hline
    $20$ & 15 & 22.0 & 32\% \\ \hline
    $20$ & 20 & 29.2& 32\% \\ \hline
  \end{tabular}
  \caption{Comparison between Yukawa couplings with and without field dependence. Note that, in the former scenario, the coupling does not flow and is equal to its UV value. The flow equations were solved with a UV cutoff $\Lambda=1\GeV$ and $\Delta g = (g_{0,k=0}-g_{0,k=\Lambda})/g_{0,k=\Lambda}$.}
\end{table}

\section{Summary}

In this paper, we raised the question of determining the field dependence of the Yukawa coupling in the three flavor quark--meson model. Our main motivation was to clarify the problem of consistency between chiral symmetry and the Yukawa term. As opposed to two flavors, in case of three flavors, the existence of non-Yukawa-like interactions and their generation in the infrared scales of the quantum effective action prevents a straightforward generalization of the two flavor approach \cite{pawlowski14}. That is to say, one cannot consistently work with, e.g., a scalar chiral condensate and associate the complete field dependence of the quark--quark--meson vertex with that of the Yukawa coupling itself. One needs great care to project out those operators that are not included in the ansatz of the effective action, which, by construction, needs to respect chiral symmetry. We believe that such a consistent approach was missing in the literature so far.

We have explicitly calculated the renormalization group flow of the field-dependent Yukawa coupling separately in the symmetric and broken phases. We have showed that at the order of dimension 6 operators new type of terms arise and gave prescription for how to project them out as required by consistency. Motivated by the very definition of the flow of the field-dependent Yukawa coupling, for the sake of a complete treatment of  dimension 6 operators, we have determined the symmetric phase flows when these new (nonrenormalizable) interactions are also included in the ansatz of the quantum effective action ($\Gamma_k$). 

Numerics showed that the field dependence of the Yukawa interaction is indeed important for physical parametrizations of the model; therefore, it would be very important to see what effects the current approach  has from a phenomenological point of view. One is typically interested in mapping the details of the chiral phase transition at finite temperature and/or density, such as the transition point, mass spectrum, interaction strengths, etc. To do so one also needs to include t' Hooft's determinant term into the system describing the $U_A(1)$ breaking, and it would be of particular interest to check the interplay between the anomaly coupling and that of the Yukawa interaction. These directions represent future studies to be reported elsewhere.

\section*{Acknowledgements}

The authors thank Zs. Szép for useful discussions related to the perturbative renormalization of the Yukawa coupling. This research was supported by the Hungarian National Research, Development and Innovation Fund under Projects No. PD127982 and K104292. The work of G. F. was also supported by the János Bolyai Research Scholarship of the Hungarian Academy of Sciences and by the ÚNKP-20-5 New National Excellence Program of the Ministry for Innovation and Technology from the source of the National Research, Development and Innovation Fund.

\makeatletter
\@addtoreset{equation}{section}
\makeatother 
\renewcommand{\theequation}{A\arabic{equation}} 

\appendix

\section{U(3) algebra identities}

The $U(3)$ algebra is spanned by the $3\times 3$ $T_i$ $(i=0,...8)$ generators, which satisfy $\Tr(T_iT_j)=\delta_{ij}/2$, and a product of two generators lie in the Lie algebra,
\bea
\label{Eq:appTT2}
T_iT_j = \frac12(d_{ijk}+if_{ijk})T_k.
\eea
Associativity of matrix multiplication leads to the following identities:
\begin{subequations}
\label{Eq:jacobi}
\bea
0&=&f_{ilm}f_{mjk}+f_{jlm}f_{imk}+f_{klm}f_{ijm}, \\
0&=&f_{ilm}d_{mjk}+f_{jlm}d_{imk}+f_{klm}d_{ijm}, \\
0&=&f_{ijm}f_{klm}-d_{ikm}d_{jlm}+d_{jlm}d_{ilm},
\eea
\end{subequations}
the first one known as the Jacobi identity. Using that $T_iT_i$ is a Casimir operator, working in the adjoint representation one easily shows that 
$f_{ijk}f_{ljk}=3\delta_{il}(1-\delta_{i0}\delta_{l0})$. Using this identity as a starting point, one derives the following two and threefold sums using (\ref{Eq:jacobi}):
\begin{subequations}
\bea
d_{ijm}d_{kjm}&=&3\delta_{ik}+3\delta_{i0}\delta_{k0}, \\
f_{lni}f_{ikm}f_{mjl}&=&-\frac{3}{2}f_{nkj}, \\
d_{mik}d_{knj}f_{jlm}&=&\frac{3}{2}f_{inl}, \\
f_{inl}f_{ljm}d_{mki}&=&\sqrt{\frac{3}{2}} (\delta_{n0}\delta_{jk}+\delta_{j0}\delta_{nk}-\delta_{k0}\delta_{nj}),\nonumber\\
&&-\frac{3}{2}d_{njk} \\
d_{ikl}d_{lnm}d_{mji}&=&\sqrt{\frac32} (\delta_{n0}\delta_{jk}+\delta_{j0}\delta_{nk}+\delta_{k0}\delta_{nj}) \nonumber\\
&&+\frac32 d_{njk}.
\eea
\end{subequations}
Furthermore, the following identities for fourfold sums are useful for present calculations:
\begin{subequations}
\bea
d_{iaj}d_{jbk}d_{kcl}d_{ldi}&=& \frac34 (d_{abm}d_{mcd}+d_{adm}d_{mcb})-\frac34 d_{acm}d_{mbd}\nonumber\\
&+&\frac12(\delta_{ab}\delta_{cd}+\delta_{ac}\delta_{bd}+\delta_{ad}\delta_{bc})\nonumber\\
&&\hspace{-1.2cm}+\frac12\sqrt{\frac32}(\delta_{a0}d_{bcd}+\delta_{b0}d_{acd} +\delta_{c0}d_{abd}+\delta_{d0}d_{abc}), \nonumber\\ \\
&&\hspace{-2cm}f_{iaj}d_{jbk}d_{kcl}d_{ldi}+f_{iaj}d_{jbk}d_{kdl}d_{lci}=f_{abj}d_{ijk}d_{kcl}d_{ldi},\nonumber\\ \\
&&\hspace{-2cm}f_{iaj}d_{jck}d_{kdl}d_{lbi}+f_{iaj}d_{jdk}d_{kcl}d_{lbi}=f_{abj}d_{ijk}d_{kcl}d_{ldi}\nonumber\\
&&\hspace{-0.4cm}+2f_{bdj}f_{lak}f_{kjm}d_{mcl}+2f_{bcj}f_{ljk}f_{kam}d_{mdl}.\nonumber\\
\eea
\end{subequations}

\renewcommand{\theequation}{B\arabic{equation}} 

\section{Invariant tensors and two-point functions}

First, we recall that 
\begin{subequations}
\bea
\rho&=& \Tr (M^\dagger M), \\
\tau&=& \Tr (M^\dagger M - \Tr(M^\dagger M)/3)^2,
\eea
\end{subequations}
where $M=(s_i+i\pi_i)T_i$. In terms of $s_i$ and $\pi_i$, they read
\begin{subequations}
\bea
\rho&=&\frac12(s_is_i+\pi_i\pi_i), \\
\tau&=&\frac{1}{24}(s_is_js_ks_l+\pi_i\pi_j\pi_k\pi_l)D_{ijkl}\nonumber\\
&+&s_is_j\pi_k\pi_l(\tilde{D}_{ij,kl}-D_{ijkl}/4) \nonumber\\
&-&\frac{1}{12}(s_is_i+\pi_i\pi_i)^2,
\eea
\end{subequations}
where $D_{ijkl}=d_{ijm}d_{klm}+d_{ikm}d_{jlm}+d_{ilm}d_{jkm}$ and $\tilde{D}_{ij,kl}=d_{ijm}d_{klm}$. The relevant derivatives of $\rho$ and $\tau$ are
\begin{subequations}
\bea
\frac{\partial \rho}{\partial s_i} &=& s_i, \quad \hspace{0.4cm} \frac{\partial \rho}{\partial \pi_i} = \pi_i, \\
\frac{\partial^2 \rho}{\partial s_is_j}&=&\delta_{ij}, \quad \frac{\partial^2 \rho}{\partial \pi_i\pi_j}=\delta_{ij},
\eea
\end{subequations}
\begin{subequations}
\bea
\frac{\partial \tau}{\partial s_i}&=&-\frac23\rho s_i+ \frac16 s_as_bs_c D_{abci} \nonumber\\
&+&\frac12s_a\pi_c\pi_d(4\tilde{D}_{ai,cd}-D_{aicd}), \\
\frac{\partial \tau}{\partial \pi_i}&=& -\frac23\rho \pi_i + \frac16 \pi_a\pi_b\pi_c D_{abci} \nonumber\\
&+&\frac12\pi_as_cs_d(4\tilde{D}_{ai,cd}-D_{aicd}),\\
\frac{\partial^2 \tau }{\partial s_i \partial s_j}&=&-\frac23 \rho \delta_{ij}-\frac23 s_i s_j + \frac12s_as_bD_{abij}\nonumber\\
&+&\frac12 \pi_c\pi_d (4\tilde{D}_{ij,cd}-D_{ijcd}), \\
\frac{\partial^2 \tau }{\partial \pi_i \partial \pi_j}&=&-\frac23 \rho \delta_{ij}-\frac23 \pi_i \pi_j +\frac12\pi_a\pi_bD_{abij}\nonumber\\
&+&\frac12 s_cs_d (4\tilde{D}_{ij,cd}-D_{ijcd}),  \\
\frac{\partial^2 \tau }{\partial \pi_i \partial s_j}&=&-\frac23\pi_i s_j + s_a\pi_c (4\tilde{D}_{aj,ci}-D_{ajci}).
\eea
\end{subequations}
In case of the broken phase calculation, we are working in a background field $M=s_0T_0+s_8T_8$; thus, the following formulas need to be used. First, we list the invariants,
\begin{subequations}
\bea
\rho|_{s_0,s_8}&=&\frac12 (s_0^2+s_8^2), \\
\tau|_{s_0,s_8}&=&\frac{1}{24} s_8^2 (8 s_0^2 - 4 \sqrt{2} s_0 s_8 + s_8^2).
\eea
\end{subequations}
Then, the nonzero first derivatives are
\begin{subequations}
\bea
\frac{\partial \rho}{\partial s_0}\bigg|_{s_0,s_8}&=&s_0, \qquad \frac{\partial \rho}{\partial s_8}\bigg|_{s_0,s_8}=s_8, \\
\frac{\partial \tau}{\partial s_0}\bigg|_{s_0,s_8}&=&s_8^2\Big(\frac{2s_0}{3}-\frac{s_8}{3\sqrt{2}}\Big), \\
\frac{\partial \tau}{\partial s_8}\bigg|_{s_0,s_8}&=&s_8\Big(\frac{2s_0^2}{3}-\frac{s_0s_8}{\sqrt{2}}+\frac{s_8^2}{6}\Big). 
\eea
\end{subequations}
As shown above, the second derivatives of $\rho$ are equal to the unit matrix, while that of $\tau$ are the following:
\begin{widetext}
\begin{subequations}
\bea
\frac{\partial^2 \tau}{\partial s_i s_j}\bigg|_{s_0,s_8}&=&
\begin{cases}
\frac{2}{3}s_8^2,  \hspace{4.5cm} \ife \hspace{0.1cm} i=j=0\\
-\frac{s_8^2}{\sqrt{2}}+\frac{4}{3}s_0s_8,  \hspace{3.0cm} \ife \hspace{0.1cm} i=0,\hspace{0.1cm} j=8 \hspace{0.1cm} \orr \hspace{0.12cm} i=8,\hspace{0.1cm} j=0\\
\frac{2}{3}s_0^2+\frac{s_8^2}{2}-\sqrt{2}s_0s_8,  \hspace{2.1cm} \ife \hspace{0.12cm} i=j=8\\
\frac{2}{3}s_0^2+\frac{s_8^2}{6}+\sqrt{2}s_0s_8, \hspace{2.1cm} \ife \hspace{0.12cm} i=j=1,2,3\\
\frac{2}{3}s_0^2+\frac{s_8^2}{6}-\frac{1}{\sqrt{2}}s_0s_8, \hspace{2.15cm} \ife \hspace{0.1cm} i=j=4,5,6,7\\
0, \hspace{4.9cm}  \els\\
\end{cases}\\
\frac{\partial^2 \tau}{\partial \pi_i \pi_j}\bigg|_{s_0,s_8}&=&
\begin{cases}
0, \hspace{4.9cm} \ife \hspace{0.1cm} i=j=0\\
-\frac{s_8^2}{3\sqrt{2}}+\frac{2}{3}s_0s_8, \hspace{2.9cm} \ife \hspace{0.1cm} i=0,\hspace{0.1cm} j=8 \hspace{0.1cm} \orr \hspace{0.1cm} i=8,\hspace{0.1cm} j=0\\
\frac{s_8^2}{6}-\frac{\sqrt{2}}{3}s_0s_8, \hspace{3.18cm} \ife \hspace{0.1cm} i=j=8\\
-\frac{s_8^2}{6}+\frac{\sqrt{2}}{3}s_0s_8, \hspace{2.9cm} \ife \hspace{0.1cm} i=j=1,2,3\\
\frac{5}{6}s_8^2-\frac{1}{3\sqrt{2}}s_0s_8 \hspace{3cm} \ife \hspace{0.1cm} i=j=4,5,6,7\\
0, \hspace{4.9cm}  \els\\
\end{cases}.
\eea
\end{subequations}
The masses of the scalar and pseudoscalar mesons of the ansatz (\ref{Eq:ansatz}) in a purely bosonic background read
\begin{subequations}
\bea
(m_{s,k}^2)_{ij}= \frac{\partial^2 V_k}{\partial s_i \partial s_j}&=&\delta_{ij}\Big(U_k'(\rho)+ \tau C_k'(\rho)\Big)+\frac{\partial^2 \tau}{\partial s_i\partial s_j}C_k(\rho)\nonumber\\
&+&\frac{\partial \rho}{\partial s_i}\frac{\partial \rho}{\partial s_j}\Big(U_k''(\rho)+\tau C_k''(\rho)\Big)+\Big(\frac{\partial \rho}{\partial s_i}\frac{\partial \tau}{\partial s_j}+\frac{\partial \rho}{\partial s_j}\frac{\partial \tau}{\partial s_i}\Big)C_k'(\rho),\\
(m_{\pi,k}^2)_{ij}=\frac{\partial^2 V_k}{\partial \pi_i \partial \pi_j}&=&\delta_{ij}\Big(U_k'(\rho)+ \tau C_k'(\rho)\Big) +\frac{\partial^2 \tau}{\partial \pi_i\partial \pi_j}C_k(\rho)\nonumber\\
&+&\frac{\partial \rho}{\partial \pi_i}\frac{\partial \rho}{\partial \pi_j}\Big(U_k''(\rho)+\tau C_k''(\rho)\Big)+\Big(\frac{\partial \rho}{\partial \pi_i}\frac{\partial \tau}{\partial \pi_j}+\frac{\partial \rho}{\partial \pi_j}\frac{\partial \tau}{\partial \pi_i}\Big)C_k'(\rho), \\
(m_{s\pi,k}^2)_{ij}=\frac{\partial^2 V_k}{\partial s_i \partial \pi_j}&=&\frac{\partial^2 \tau}{\partial s_i\partial s_j}C_k(\rho)+\frac{\partial \rho}{\partial s_i}\frac{\partial \rho}{\partial \pi_j}\Big(U_k''(\rho)+\tau C_k''(\rho)\Big)+\Big(\frac{\partial \rho}{\partial s_i}\frac{\partial \tau}{\partial \pi_j}+\frac{\partial \rho}{\partial \pi_j}\frac{\partial \tau}{\partial s_i}\Big)C_k'(\rho).
\eea
\end{subequations}
If the fermions also have nonzero expectation values, then these masses get corrected by
\begin{subequations}
\label{Eq:deltamass53}
\bea
\Delta (m_{s,k}^2)_{ij} &=& \bar{q}\Big[g_k'(\rho)\Big(\frac{\partial \rho}{\partial s_i}\frac{\partial M_5}{\partial s_j}+\frac{\partial \rho}{\partial s_j}\frac{\partial M_5}{\partial s_i}\Big)+g_k'(\rho)\delta_{ij}M_5+g_k''(\rho)\frac{\partial \rho}{\partial s_i}\frac{\partial \rho}{\partial s_j}M_5\Big]q \nonumber\\
&+&g_{53}\bar{q} \big(\{\{T_i,T_j\},M_5\}+T_iM_5^\dagger T_j + T_j M_5^\dagger T_i\big) q, \\
\Delta (m_{\pi,k}^2)_{ij} &=& \bar{q}\Big[g_k'(\rho)\Big(\frac{\partial \rho}{\partial \pi_i}\frac{\partial M_5}{\partial \pi_j}+\frac{\partial \rho}{\partial \pi_j}\frac{\partial M_5}{\partial \pi_i}\Big)+g_k'(\rho)\delta_{ij}M_5+g_k''(\rho)\frac{\partial \rho}{\partial \pi_i}\frac{\partial \rho}{\partial \pi_j}M_5\Big]q \nonumber\\
&+&g_{53}\bar{q} \big(\big\{M_5,\{T_i,T_j\}\big\}-T_iM_5^\dagger T_j - T_j M_5^\dagger T_i\big) q, \\
\Delta(m_{s\pi,k}^2)_{ij} &=& \bar{q}\Big[g_k'(\rho)\Big(\frac{\partial \rho}{\partial s_i}\frac{\partial M_5}{\partial \pi_j}+\frac{\partial \rho}{\partial \pi_j}\frac{\partial M_5}{\partial s_i}\Big)+g_k''(\rho)\frac{\partial \rho}{\partial s_i}\frac{\partial \rho}{\partial \pi_j}M_5\Big]q \nonumber\\
&+&ig_{53}\bar{q} \gamma_5 \big( \big[M_5,[T_i,T_j]\big]+T_i M_5^\dagger T_j + T_j M_5^\dagger T_i\big) q.
\eea
\end{subequations}
where we also included the effect of the $\sim\! \bar{q} M_5 M_5^\dagger M_5 q$ operator with the coupling constant $g_{53}$. Using the formulas above, in a purely bosonic background, which is defined by $M=s_0T_0+s_8T_8$, the two-point functions of the ansatz (\ref{Eq:ansatz}) read
\begin{subequations}
\label{Eq:gammas}
\bea
\Gamma^{(2)00}_{k,s}(p)&=&p^2+U_k' + \frac{1}{24} \Big(8 (2 C_k + 5 C_k' s_0^2 + C_k'' s_0^4) s_8^2 - 4 \sqrt{2} s_0 (3 C_k' + C_k'' s_0^2) s_8^3 + (C_k' + C_k'' s_0^2) s_8^4 + 24 s_0^2 U_k''\Big), \nonumber\\ \\
\Gamma^{(2)08}_{k,s}(p)&=&\frac{1}{24} s_8 \Big(-12 \sqrt{2} (C_k + C_k' s_0^2) s_8 + 4 s_0 (5 C_k' + 2 C_k'' s_0^2) s_8^2 - 4 \sqrt{2} (C_k' + C_k'' s_0^2) s_8^3 + C_k'' s_0 s_8^4 \nonumber\\
&&\hspace{9.66cm}+ 8 s_0 (4 C_k + 2 C_k' s_0^2 + 3 U_k'')\Big), \\
\Gamma^{(2)88}_{k,s}(p)&=& p^2+U_k' +\frac{1}{24} \Big(8 s_0^2 (2 C_k + 5 C_k' s_8^2 + C_k'' s_8^4) - 4 \sqrt{2} s_0 s_8 (6 C_k + 7 C_k' s_8^2 + C_k'' s_8^4) \nonumber\\
&&\hspace{8.4cm}+ s_8^2 (12 C_k + 9 C_k' s_8^2 + C_k'' s_8^4 + 24 U_k'')\Big),\\
\Gamma^{(2)00}_{k,\pi}(p)&=&p^2+U_k'+\frac{1}{24} C_k' s_8^2 (8 s_0^2 - 4 \sqrt{2} s_0 s_8 + s_8^2), \\
\Gamma^{(2)08}_{k,\pi}(p)&=&-\frac16 C_k s_8 (-4 s_0 + \sqrt{2} s_8),\\
\Gamma^{(2)88}_{k,\pi}(p)&=&p^2+U_k'+\frac{1}{24}s_8 \Big(4 C_k (-2 \sqrt{2} s_0 + s_8) + C_k' s_8 (8 s_0^2 - 4 \sqrt{2} s_0 s_8 + s_8^2)\Big),
\eea
\end{subequations}
and furthermore,
\begin{subequations}
\label{Eq:gammas2}
\bea
\Gamma^{(2)11}_{k,s}(p)&=&p^2 +U_k' + \frac{1}{24} \Big( 4 C_k s_8^2  + C_k' s_8^4  + 8 s_0^2 (2 C_k + C_k' s_8^2) - 2 \sqrt{2} s_0 s_8 (-12 C_k  + 2 C_k' s_8^2) \Big),\\
\Gamma^{(2)44}_{k,s}(p)&=&p^2+U_k' +\frac{1}{24} \Big(4 C_k s_8^2 + C_k' s_8^4 + 8 s_0^2 (2 C_k + C_k' s_8^2) - 2 \sqrt{2} s_0 s_8 (6 C_k + 2 C_k' s_8^2) \Big),\\
\Gamma^{(2)11}_{k,\pi}(p)&=&p^2+U_k' +\frac{1}{24} s_8 \Big(4 C_k(2 \sqrt{2} s_0 -  s_8) + C_k' s_8 (8 s_0^2 - 4 \sqrt{2} s_0 s_8 + s_8^2)\Big),\\
\Gamma^{(2)44}_{k,\pi}(p)&=&p^2+U_k'+\frac{1}{24} s_8 \Big(4 C_k (-\sqrt{2} s_0 + 5 s_8) + C_k' s_8 (8 s_0^2 - 4 \sqrt{2} s_0 s_8 + s_8^2)\Big),
\eea
\end{subequations}
where $\Gamma^{(2)11}_{k,s/\pi}(p)=\Gamma^{(2)22}_{k,s/\pi}(p)=\Gamma^{(2)33}_{k,s/\pi}(p)$, and $\Gamma^{(2)44}_{k,s/\pi}(p)=\Gamma^{(2)55}_{k,s/\pi}(p)=\Gamma^{(2)66}_{k,s/\pi}(p)=\Gamma^{(2)77}_{k,s/\pi}(p)$.

\renewcommand{\theequation}{C\arabic{equation}} 

\section{Broken phase flows of $g_0$ and $g_1$}

Using the expansion (\ref{Eq:gexpan}) and the flow equation (\ref{Eq:npflow}), we get
\bea
\label{Eq:brokeng0}
\partial_k g_{0,k}&=&\int_p \tilde{\partial}_k\Bigg[\frac{3g_{0,k}^3/2}{(p_R^2+g_{0,k}^2\rho_0/3)(p_R^2+U_k'(\rho_0))}-\frac{4g_{0,k}^3/3}{(p_R^2+g_{0,k}^2\rho_0/3)(p_R^2+U_k'(\rho_0)+4C_k(\rho_0)\rho_0/3)}\nonumber\\
&&\hspace{1cm}-\frac{g_{0,k}^3/6+2\rho_0 g_{0,k}^2g_{1,k}/3+2\rho_0^2g_{0,k} g_{1,k}^2/3}{(p_R^2+g_{0,k}^2\rho_0/3)(p_R^2+U_k'(\rho_0)+2\rho_0 U_k''(\rho_0))} +\frac{9g_{1,k}/2}{p_R^2+U_k'(\rho_0)}+\frac{4g_{1,k}}{p_R^2+U_k'(\rho_0)+4C_k(\rho_0)\rho_0/3}\nonumber\\
&&\hspace{1cm}+\frac{3g_{1,k}/2}{p_R^2+U_k'(\rho_0)+2\rho_0 U_k''(\rho_0)}\Bigg],\\
\partial_k g_{1,k}&=&\int_p \tilde{\partial}_k \frac{d}{d\rho}\Bigg[\frac{3g_k^3(\rho)/2}{(p_R^2+g_k^2(\rho)\rho/3)(p_R^2+U_k'(\rho))}-\frac{4g_k^3(\rho)/3}{(p_R^2+g_k^2(\rho)\rho/3)(p_R^2+U_k'(\rho)+4C_k(\rho)\rho/3)}\nonumber\\
\label{Eq:brokeng1}
&&\hspace{1cm}-\frac{g_k^3(\rho)/6+2\rho g_k^2(\rho)g_k'(\rho)/3+2\rho^2g_k(\rho) g_k'^2(\rho)/3}{(p_R^2+g_k^2(\rho)\rho/3)(p_R^2+U_k'(\rho)+2\rho U_k''(\rho))} +\frac{9g_k'(\rho)/2}{p_R^2+U_k'(\rho)}+\frac{4g_k'(\rho)}{p_R^2+U_k'(\rho)+4C_k(\rho)\rho/3}\nonumber\\
&&\hspace{1cm}+\frac{3g_k'(\rho)/2}{p_R^2+U_k'(\rho)+2\rho U_k''(\rho)}\Bigg]_{\rho=\rho_0},
\eea
where $g_k(\rho)=g_{0,k}+g_{1,k}(\rho-\rho_0)$.

\end{widetext}

\end{document}